# The Science Wars: a pox on both their houses

Stephen Boughn*

Departments of Physics and Astronomy, Haverford College, Haverford, PA 19041

**The Science Wars**

According to physicist David Mermin, the *science wars* was a series of exchanges between scientists and "sociologists, historians, and literary critics" whom the scientists thought to be "ludicrously ignorant of science, making all kinds of nonsensical pronouncements. The other side dismissed these charges as naive, ill informed and self-serving". [1]   Another perspective was expressed by sociologist Harry Collins who wrote that the science wars "began in the early 1990s with attacks by natural scientists or ex-natural scientists who had assumed the role of spokespersons for science. The subject of the attacks was the analysis of science coming out of literary studies and the social sciences." [1]

My first introduction to the conflict occurred in the early 1990s at a lunch conversation with a (postmodern?) historian friend of mine.  I had just responded to a question about the Big Bang, which elicited the following response: "So I see, the Big Bang Theory is just the latest in a long line of creation myths."  In retrospect, I think my friend was just yanking my chain but, of course, I felt compelled to reply and said something like, "There are indeed various big bang theories but I'm an experimental physicist and observational astronomer and, for me, the big bang is more of an observation than a theory.  We observe that distant galaxies are all rushing away from us with speeds such that one infers they were all at the same place at the same time, some 10 billion years ago."  Of course, this wasn't convincing to my friend and I proceeded to present more detailed arguments, which only served to acknowledge that my chain had been duly yanked.

As a consequence of that encounter, I read a compilation of essays on postmodernism but was able to comprehend very little of what I read.  Some of the material seemed silly to me but I am quick to admit that I lacked sufficient knowledge, not to mention the vocabulary, to judge the essays. After all, I'm sure that a postmodernist presented with a volume of essays on quantum field theory would have understood even less.

---

* sboughn@haverford.edu

Some postmodernists suggested that scientific theories were social constructs which raised the ire of some scientists, especially physicists, who saw such theories as representations of objective reality. From my limited perspective this was the conflict that initiated the science wars. Some have attributed the social constructionist gauntlet to their jealousy of the sciences. I rather think it was a yank on the chain of (some) arrogant physicists. I'm restricting my essay to the social constructionist/physics conflict because I am, after all, a physicist. According to some social constructionists: *the fundamental theories of physics are simply social constructs.* The knee-jerk response of physicists to this declaration is invariably something like: *the fundamental theories of physics are good approximations to the true laws that govern the objective reality we refer to as nature.* Even though there are many more nuanced aspects of the science wars, I believe that this simple exchange characterizes the essence of the conflict.

In 1995, biologist Paul Gross and mathematician Normal Levitt authored a book entitled, *Higher Superstition: The Academic Left and Its Quarrels with Science* [2]. "They asserted that anti-intellectual sentiment in liberal arts departments (especially English departments) caused the increase of deconstructionist thought, which eventually resulted in a deconstructionist critique of science."[3] This book motivated physicist Alan Sokal to submit a 1996 article to *Social Text,* an academic journal of cultural studies. The article , "Transgressing the Boundaries: Towards a Transformative Hermeneutics of Quantum Gravity"[4], was published without review and soon thereafter was revealed by Sokal to be a hoax. Sokal's purpose was to demonstrate the intellectual laziness of the American academic left. The Sokal Affair, as it has been dubbed, brought the science wars to the forefront of academia and into the mainstream media. I confess that I've not read Sokal's article nor his follow up book; however, I did attend his Princeton physics department colloquium on the topic. While I was amused, I didn't find anything he presented to be very interesting. Perhaps he demonstrated the laziness of some editors at *Social*

*Text* but I've certainly read questionable papers in physics journals that do use peer review. Because Sokal was a theoretical physicist and the topic involved quantum gravity, perhaps the editors didn't feel competent to challenge him. And, who knows, maybe some of Sokal's gibberish did contain something significant to experts in the field. After all, a monkey at a typewriter may on occasion type something intelligible.

The Sokal Affair was, perhaps, the peak of the science wars, which then descended back into the obscurity of academia, or so I thought. I recently read two books that made me reconsider and led me to pen this essay. The first was the latest book by Nobel laureate James Peebles with the title *The Whole Truth: A Cosmologist's Reflection on the Search for Objective Reality* [5]. Sounds to me like a shot across the bow of the science wars.[1] The other book, *The War on Science* [6], by Shawn Otto, was not a response to the science wars but rather describes the increasing anti-science attitude that has been developing for the last 50 to 60 years. This "war" comes from many directions: political, cultural, religious, left wing, right wing, you name it. We're all familiar with recent symptoms including nuclear weapons, the dangers inherent in nuclear power stations, the resistance of teaching evolution in public schools, vaccine hesitancy, the safety of genetically modified foods, and the reluctance to accept human influence on the earth's climate. The "war on science" is certainly distinct from the "science wars"; however, Otto does place part of the blame on the postmodernism movement in universities. I rather doubt that an arcane academic dispute could have much influence on the public. Nevertheless, in the appendix to this essay I'll offer a few words on a source of anti-science sentiment.

There were several attempts to bring together scientists and postmodernists to reconcile their differences, one of which gave rise to a book, *The One Culture?*, edited by chemist Jay A. Labinger and sociologist Harry Collins [7], wherein 9 scientists and 6 sociologists and historians of science contributed initial essays, then commentaries on the essays, and finally rebuttals to the commentaries. Seven of the nine scientists were physicists (including two Nobel laureates), which validated my impression that physicists felt the most aggrieved by social constructionists' proclamations.  I also perused this book and, as I expected, it confirmed my expectation that those who specialize in the sociology of science are intelligent beings with reasonable reflections on the ways in which scientists pursue their professions.  Also, as one might expect, a great deal of the science wars arose because of differences in the vocabularies of the two warring parties, which the book's format did much to alleviate.   On the other hand, I have never had more than a passing interest in sociology nor did I take a sociology course at university (homo sapiens are a much too complicated system for me to understand); therefore, I didn't learn much from these essays. While there seemed to be some reconciliation, I sensed that there remained tension between the two groups.  In particular, the responses of the scientists made clear to me that they have a deeply ingrained, absolute "belief" (and I use this word purposely) in an objective reality and the true laws that govern that reality.  I now move on to my take on the source of this tension and why I chose the cheeky subtitle for this essay.

---

[1] Jim Peebles has been my teacher, mentor, colleague, and friend for over 50 years. I have yet to ask him about the motivation for the title of his book but I must do so. ☺

**A Pox On Both Their Houses**

Let me begin by posing what, as mentioned above, I consider to be the crux of the science wars.  Admitting that the source of the conflict is more subtle than the interaction I had with my historian friend all those years ago, let's start there. Suppose a (perhaps radical) social constructionist confronts a physicist with the declaration:

Postmodernist:
"The fundamental laws of physics are simply *social constructs*."

The physicist immediately replies:
 "No. The fundamental theories of physics are our best approximations to the *true laws* that govern the *objective reality* we call nature."

Both of these declarations make grammatical sense and, at first glance, seem to be straightforward statements that can be judged according to their merits.  I certainly proceeded to argue with my historian friend without clarification. But not so fast.

Without further explaination, such an argument makes about as much sense as a squabble about how many angels can dance on the head of a pin. So let's seek further clarification. Because I'm a physicist, I'll first interrogate the social constructionist with the following:

"Just what do you mean by *social constructs*? If you mean that the mathematical models expressing the fundamental theories of physics were written down by physicists who were members of particular cultures and societies then, of course, your statement is accurate; however, it is also nearly devoid of content. The same could be said about anything we do or say. On the other hand, if you mean that physicists who are members of different cultures and societies will naturally subscribe to different fundamental laws of physics then, of course, that's nonsense. Mathematics and physics have been developed over the last four millennia by people living in vastly different cultures and societies all over the world and yet, today, they all subscribe to the same laws of physics."

My social constructionist friend might respond with something like:

"This is undoubtedly due to the communication between these different societies that has developed over these same millennia. *IF* two advanced societies were never in communication with each other then they quite likely would have arrived at vastly different fundamental laws of physics."

My response:

"Now you've introduced *counterfactuals* into your argument. I've never found counterfactual arguments at all convincing so I see no reason to continue our conversation."

With one last attempt, my social constructionist friend might say:

"*When* we make contact with an advanced extraterrestrial civilization, then it is quite likely we will find that they subscribe to completely different fundamental laws of physics."

And my response is:

"I admit that you've successfully removed the counterfactual from your argument but until we do communicate with an advanced extraterrestrial civilization, I still don't see any purpose for continuing our discussion."

Okay, I think I've made a cogent point of view for the physicists. Now, putting on my social constructionist hat, let me interrogate the physicist:

"Just what do you mean by *true laws* and *objective reality*? Surely you don't mean there is a perfect model of reality in the realm of the gods on Mt. Olympus accompanied by a set of god given rules that govern that reality?"

My physicist friend might say something like:

"Of course not. Objective reality just *is* as are the laws that govern it."

Then I would say:

"I'm sorry but that doesn't help. You must be more specific about what you mean. Perhaps you can give me a proof of the existence of objective reality."

The physicist responds:

"Absolutely. Our mathematically models, as you call them, are very precise and they accurately predict nearly all that we observe. How else can we explain that this is the case except by postulating an objective reality *out there* and recognizing that our models are accurate approximations to the laws that govern this reality."

My response:

"Hmm…This response reminds me of the cosmological proof for the existence of God, that is, how else are we to explain how our wonderful universe came into being? It must have been created and that creator is who we call 'God'. However, then one might ask 'what created God?' and so on ad infinitum. A theologian might answer, 'God is the primary cause of everything. It is not necessary to go further', which to me simply means, 'I don't want to continue talking about it'. For a physicist like me, Ockham's razor cautions us not to invent an explanation whose sole purpose is to explain something else but is otherwise of no use. That is, why not just declare that our wonderful universe exists and stop there.

What does this have to do with objective reality? Well, my physicist friend, you just explained the accuracy of our physical theories by claiming that they are good approximations to the laws that govern objective reality. But then the only purpose of postulating 'objective reality', is to explain the usefulness of our physical theories and nothing else. Why not just note that these mathematical models are very good and stop there as Ockham's razor advises us."

My physicist friend tries one last tactic:

"I'm tempted to declare that *when* we communicate with an advanced extraterrestrial civilization, they will profess a similar belief in objective reality and true laws that govern it. However, from your response to our postmodern friend, I realized that won't satisfy you.  So what if I claim that if and when there is such contact, we can be assured that extraterrestrial predictions of physical phenomena will be the same as ours.  I know this because we observe that matter and radiation in the far reaches of the universe behave the same way as in our neighborhood."

My response:

"A very good point and one that you should make in arguments with your postmodernist friends.  However, notice that you forewent any discussion of extraterrestrial laws of objective reality.  (Had you not, I would have accused you of engaging in a discussion equivalent to how many extraterrestrial angels can dance on the head of an extraterrestrial pin ☺.) So why even bring up objective reality?"

**Final Remarks**

I've compared the "belief" (there's that word again) in an objective reality with a belief in God and I fully accept this comparison.  It's not necessarily bad to have beliefs but beliefs should not provide the foundation for arguments such as those of the science wars.  I find it ironic that physicists base their argument on metaphysical concepts.  In fact, one might argue that the very *belief* in true laws of objective reality is a social construct, which only heightens the irony. This is the fault that I find with most of the physicist combatants.  Of the contributors in "The One Culture?" text of essays [7], physicists Alan Sokal and Steven Weinberg made the strongest arguments for the relevance of a true physical reality; however, even they hinted that this reality is more a belief than a demonstrable description of our world.  Sokal remarked that the notion of *truth* itself can only be grasped intuitively while Weinberg seems to acknowledge that the meaning of *reality* may never be pinned down but that doesn't preclude us from the common use of the term, what he calls "naïve realism".

So do I believe in either "God" or "objective reality".  Certainly not in the former.  With regard to the latter, I would say that I believe that there is a *real world* out there that I experience every morning when I wake up.  However, I strongly advocate that in discussions of physics topics the words "real" and "reality" should be avoided at all costs.[2]  With regard to the two questions, "Does God exist?" and "Does objective reality exist?", I am agnostic.  These questions can never be answered; therefore, why even pose them.  So what possible use is there in a belief in objective reality?  Some (most?) physicists might say that such a belief is what motivates them to search for (I would say create) the mathematical models that are so useful in describing the natural world.  I acknowledge that this is undoubtedly the case but the source of scientific creativity is beyond my poor powers and I have nothing to offer here.  I suspect that sociologists and psychologists may be better equipped to discuss this topic.

      I have been quite critical of the physicist side of the science wars but I don't have the background to say much about the postmodernist side.  As I've mentioned above, I'm quite willing to accept that some of the deliberations of the social constructionists might well be of interest to many sociologists even if they are of little interest to me.[3]  I think I've made my objections clear above but here's a recap.  If social scientists and postmodernists wish to converse with scientists, physicists in particular, they must endeavor to speak in a way that physicists can understand.  Saying "the laws of physics are social constructs" doesn't fit this bill.  The meaning of this thesis must be made explicit as well as its demonstrable consequences made clear, that is, what follows from it.  I initially assumed that the source of the conflict was the inability of social constructionists to explain to physicists what they mean when they make provocative statements.  Of course it's possible that the science wars were provoked not by postmodernists but by physicists who began paying attention to the social construction literature without the background to fully understand what's being said.  Or, perhaps, the provocative statements were simply motivated by a desire of some postmodernists to jerk the chains of physicists.

      Ergo, a pox on both their houses.

---

[2] I wrote a short essay expressing this view, "Against Reality in Physics". [8]

[3] I didn't even find Thomas Kuhn's book, *The Structure of Scientific Revolution,* to be very interesting; although, I've just begun reading *The Last Writings of Thomas Kuhn.*  Maybe that will help.

**Appendix: The War on Science**

I think the term, "the war on science", probably arose from the title of Shawn Otto's 2016 book, *The War on Science: who's waging it; why it matters; what we can do about it.* I found this 500 page book to be an interesting account of the rising anti-science sentiment in the last half century. He points out many possible sources of this trend and offers many reasonable responses to it. As I mentioned above, Otto places part of the blame on the postmodernist movements in universities, a claim with which I take issue. On the other hand, I do think there is a connection between "the science wars" and "the war on science" as I'll explain shortly. First, let me tell you what the terms "science" and "scientist" mean to me. I make no attempt to trace the history of how these terms arose, a topic that I'm sure would require thousands of pages of historical, sociological, and "scientific" investigation.

Let me begin with a personal account of how I became a scientist. When I was about 5 or 6 years old I became enthralled with science, which is to say that I loved astronomy and dinosaurs. This interest continued and expanded for the next few years[4]. Sometime later I heard about the "scientific method" and was thrilled that I would soon learn about this mystical code and fulfill my dream of becoming a scientist. The only problem was when I finally learned the details of the scientific method I was very disappointed. It just seemed like a string of commonsense statements (except, perhaps, for the one about forming hypotheses, which I didn't fully understand - still don't). It seemed to me that these rules were the common sense practices that we all follow every day. I remember thinking that my dad, a welder and self-taught machinist who often designed and fabricated specialized equipment for farmers and ranchers, must surely be a scientist. In fact, it seemed that everyone I knew was a scientist. We all experience events around us and then decide what to do in order to produce a desired outcome. If it works then that reasoning becomes part of our understanding of the world; if it doesn't then we make other guesses (hypotheses) and try again.

---

[4] I began an archeological excavation of the vacant lot behind our backyard and duly recorded my findings in a notebook: a fossil squid, a gas lamp, a garter snap, etc. My dad then informed me that the lot used to be a garbage dump, which ended that enterprise. (I kept the fossil squid.)

So when a newspaper headline tells us that "Scientists make a groundbreaking discovery!", what do they mean? Just who are "scientists"? At best, the term is simply a stand in for something more descriptive that will come later in the article. I guess that's okay but it seems silly. Why not say something meaningful right away, like "Astronomers make a groundbreaking discovery!" However, I suspect that there is another, perhaps hidden, message implied by "scientists" and that is a declaration that what is about to be said is something that you will not be able to fully understand if you are not a member of that mystical cult of beings who have a monopoly on understanding the laws that govern the objective reality of the world. The implication is that a nonscientist must simply defer to the scientist on the matter.[5] This brings us right back to the argument used by physicists in the science wars: our fundamental theories are not social constructs but are our best approximations to the *true laws* that govern the *objective reality* we call nature.

So how does this claim contribute to *the war on science*? If one labels this claim a *belief* as do I, then in some sense it's occupies the same footing as a religious belief. If this is the case, then it seems entirely reasonable that not everyone need accept this belief, especially if it is in conflict of some other of one's beliefs. On the other hand, if the scientific method is simply common sense that we all follow, then who can possibly object? To be sure, the methods of a practitioner of any endeavor can be incredibly complex and the details not fully understood by most people. But that doesn't mean that one cannot make informed decisions about the subject of that endeavor. I may have little knowledge about the mechanisms in an electric automobile but I know enough to make an informed decision about buying one.

One of Shawn Otto's laments is that journalists and the public in general don't know enough science to make informed decisions about, for example, the claim that human use of fossil fuels causes the rise in global temperatures. One of his suggested responses to this situation is to increase science education of everyone, journalists in particular. I seriously doubt that this is remotely possible. I have a PhD in experimental physics but I'm sure it would take me a year of hard study to be able to critically examine the climate models that imply such a temperature increase. So what are we to do? Simply accepting conclusions of general climate models in the absence of any understanding of the matter doesn't seem right. Who are these people and why are they telling us how to live our lives? However, with a cursory account of the phenomenon along with an emphasis on the inherent uncertainty in climate models, I claim that the average citizen can achieve sufficient understanding to at least begin to consider how we might respond to the potential risks of climate change.

---

[5] It's for these reasons that when I encounter a piece that begins "Scientists have found…." I conclude that either the writer has minimal understanding of the topic or that I'm about to be a victim of some sort of propaganda. Either way, I usually stop reading (or listening).

How might this be accomplished? To begin with, the greenhouse warming effect is relatively easy to understand. Most of us have seen a glass greenhouse. The sunlight comes through the glass walls and roof of the structure and warms the interior while those same walls and roof impede convective cooling with the result that the inside temperature is higher than the outside temperature. In the case of the earth's atmosphere, not convection but radiation is the source of heat loss; however, the general idea is the same. We've known for a hundred years that in the absence of an atmospheric greenhouse effect, the average temperature of the earth would be an inhospitable 0 °F compared to the actual temperature of 57 °F. On the other hand, the carbon dioxide atmosphere of Venus is so thick that the average surface temperature is 867 °F , hotter than the melting point of lead, This was determined more than 50 years ago. The burning of fossil fuels certainly contributes to the $CO_2$ in the atmosphere but the question is by how much does this increase the global temperature. So far, this explanation is accessible even to a 7 year old. This brings us to climate models and the level of uncertainty in those models.

Current climate models indicate that the burning of fossil fuels and the concomitant increase in the $CO_2$ content of the atmosphere are largely responsible for the increase in global temperature. But how certain are we about this

12conclusion? Clearly world climate is an incredibly complex system and I would be astonished if there were not considerably uncertainties in climate models. Otto decries the American Petroleum Institute's focus on the uncertainties of climate models. On the other hand, combating this strategy with overstating the accuracy of these models is propaganda of another type. One such claim that I frequently hear is "97% of climate scientists say that observed climate warming is due to human use of fossil fuels." This may sound like a claim about the certainty of climate models but I have no idea what is means. Does it mean that 97% of climate modelers think that there is at least a 51% chance that we are responsible for climate change while only 3% of modelers think that there is less than a 50% chance that we are responsible for climate change? Or, perhaps, does it mean that 97% of climate modelers are 100% sure that we are responsible for climate change and that 3% of modelers are 100% sure that we are in no way responsible for climate change. What it certainly doesn't mean is that there is a 97% certainty that humans are the major cause of climate change. I seriously doubt that the accuracy of climate models can be so precisely estimated. Why not simply accept the fact that there is considerable uncertainty in climate models and then concentrate on the serious consequences of the observed increase in global temperature.

When I discuss climate change with my right-wing, skeptical friends, I find that the most effective argument emphasizes the uncertainty of climate models. I might say something like: "Of course climate models are quite uncertain; in fact, let's assume that there is only a 30% chance that the models correctly predict a human cause of the rise in global temperature. It might seem foolish to make draconian changes in public policy based on 30% odds. However, if that 30% prediction should come about, in 50 to 100 years from now when Florida and all US coastal cities are under water, what do you say? Oops!"

Most everyone has an intuitive understanding of probability and statistics (except maybe when it comes to national lotteries ☺). For example, I suspect that most people might predict that there is on the order of a 30% chance they will be struck by a car and possibly killed if they jaywalk across a busy city street with their eyes closed. (It seems most likely that drivers would slam on their breaks, honk their horns, and shout epithets.) That's better than a 2 to 1 chance you will safely cross the street with your eyes closed; so why not do it? Of course you would not because of the possible dire consequence.

So my conclusion is that we don't need a massive effort to educate journalists and the public on the details of science but rather recognize that all science, no matter how devilish the details, is simply based on common sense and is judged by its utility rather than on some mystical notion of truth. If the uncertainty in a prediction is hard to estimate, as in the case of climate models, so be it but one can still act on those predictions if the consequences are truly catastrophic. For me, the fact that even some of the climate models indicate that humans are responsible for catastrophic climate change is enough to motivate serious consideration of societal behavior.

Now I'm beginning to ramble so I'd better quit.

**Acknowlegements**

I realize that my views on science, physics in particular, my be quite different from the majority of physicists. I've come to them primarily from my readings of William James, Bohr, Heisenberg, Ludwig Wittgenstein, Henry Stapp, and discussions with Freeman Dyson and Marcel Reginatto. I don't claim that they would agree with all that I have expressed in this essay but think that they would have at least understood what I'm saying (except maybe Wittengenstein ☺).

**References**

[1] "Science Wars", Wikipedia.

[2] Gross, P. and N. Levitt, Norman, *Higher Superstition: The Academic Left and Its Quarrels With Science*, (Baltimore: Johns Hopkins University Press, 1995).

[3] "Sokal Affair", Wikipedia.

[4] Sokal, A., ""Transgressing the Boundaries: Toward a Transformative Hermeneutics of Quantum Gravity", *Social Text (*Duke University Press 1996).

[5] Peebles, P.J.E., *The Whole Truth: A Cosmologist's Reflection on the Searth for Objective Reality*, (Princeton University Press, Princeton, NJ, 2022).

[6] Otto, S., *The War on Science: Who's Waging It; Why It Matters; What We Can Do About It,* (Milkweed Editions, Minneapolis, Minnesota, 2016).

[7] Laginger, J., H. Collins, *The One Culture: A Conversation about Science,* (Chicago University Press, Chicago, 2001).

[8] Boughn, S., "Against 'Reality' in Physics", arXiv:1903.08673 [physics.hist-ph; quant-ph] (2019).